\documentstyle[aps,prl,preprint]{revtex}
\newcommand{\newc}{\newcommand}
\newc{\gsim}{\lower.7ex\hbox{$\;\stackrel{\textstyle>}{\sim}\;$}}
\newc{\lsim}{\lower.7ex\hbox{$\;\stackrel{\textstyle<}{\sim}\;$}}
\begin{document}
\begin{titlepage}

\baselineskip=14pt
\begin{flushright}
{\footnotesize
FERMILAB--Pub--96/410\\
UMD-PP-97-54\\
Submitted to Phys. Rev. D}
\end{flushright}
\renewcommand{\thefootnote}{\fnsymbol{footnote}}
\vspace{0.15in}
\baselineskip=24pt

\begin{center}
{\Large \bf Supersymmetric Models with Anomalous $U(1)$ Mediated Supersymmetry 
Breaking\\}
\baselineskip=14pt
\vspace{0.75cm}

\vspace{0.3cm}
{\bf R.N. Mohapatra$^{(1)}$\footnote{Electronic
address: {\tt rmohapatra@umdhep.umd.edu}} and 
 A. Riotto$^{(2)}$\footnote{Electronic address:
{\tt riotto@fnas01.fnal.gov}}}, \\
\vspace{0.4cm}
$^{(1)}${\em Department of Physics and Astronomy,
University of Maryland,\\ College Park, MD~~20742, USA}\\
\vspace{0.3cm}
$^{(2)}${\em NASA/Fermilab Astrophysics Center,\\
Fermi National Accelerator Laboratory,\\ Batavia, Illinois~60510-0500, USA}\\
\vspace{0.3cm}

\end{center}

\baselineskip=14pt

\begin{quote}
\hspace*{1em}
\begin{center}
{\bf\large Abstract}
\end{center}
\vspace{0.2cm}

We construct realistic supergravity models where supersymmetry 
breaking arises from the $D$-terms of an anomalous $U(1)$ gauge symmetry
broken at the Planck scale. Effective action for these theories at
sub-Planck energies (including higher dimensional terms in the 
superpotential) are severely restricted by the $U(1)$ symmetry and 
by the assumption they arise from an underlying renormalizable
theory at a higher scale. Phenomenological consequences of these
models are studied. It is found that they have the attractive feature
that the gaugino masses, the $A$ and $B$ terms and
the mass splittings between the like-charged 
squarks of the first two generations compared to their average 
masses can all be naturally suppressed. As a result, the
electric dipole moment of the neutron  as well as the flavor changing
neutral current effects  are predicted to be naturally small.
These models also predict the value of the $\mu$-term to be naturally
small and have the potential to qualitatively explain the
observed mass hierarchy among quarks and leptons. We then discuss examples of
high scale renormalizable theories that can justify the choice of the
the effective action from naturalness point of view.

\vspace*{8pt}


\renewcommand{\thefootnote}{\arabic{footnote}}
\addtocounter{footnote}{-2}
\end{quote}
\end{titlepage}

\baselineskip=24pt
\section{Introduction}

Supersymmetry provides ways to solve many of the puzzles of the
standard model such as the stability of the weak scale under radiative
corrections as well as the origin of the weak scale itself. Local
supersymmetry provides a promising way to include gravity within the
framework of unified theories of particle physics eventually leading the
way perhaps to a theory of everything in string models. For this very good
reason, supersymmetric extensions of the standard model (MSSM) 
have been the focus of intense theoretical activity\cite{nilles} in recent
years. Since experimental observations require supersymmetry to
be broken, it is essential to have a knowledge of
the nature and the scale of supersymmetry breaking in order to have a
complete understanding of the physical implications of these theories.
At the moment, we lack such an understanding and therefore it is important 
to explore the various ways in which supersymmetry breaking can arise and
study their consequences.

There are several hints from the study of general
class of MSSM which could perhaps be useful in trying to explore
the nature of supersymmetry breaking. 
Two particular ones that rely on the supersymmetric sector of model are: 
(i) natural suppression of flavor changing neutral
currents (FCNC) which require a high degree of degeneracy among squarks
of different flavor and (ii) stringent upper limits on the electric
dipole moment of the neutron (NEDM) which imply constraints on the
gaugino masses as well as on the $A$ and $B$ terms of MSSM\cite{dine}.
One could take the point of view that the above conclusions may be telling
us something about the nature of supersymmetry breaking. If this is true,
then it is important to isolate those SUSY breaking scenarios which 
realize the above properties in a simple manner and study their implications.

The standard way in which supersymmetry breaking is implemented in
model building is to postulate the existence of a hidden sector
where local supersymmetry is spontaneously broken and then find an
appropriate way to transmit them to the visible sector. The various classes
of models can be isolated depending on the way the SUSY breaking is
transmitted. The two popular ones widely discussed in the literature 
are: (a) Polonyi type models
where the SUSY breaking is transmitted via the gravitational interactions.
The typical scale of SUSY breaking in such models is of order of  $\sqrt{M_W M_{P\ell}}\simeq 10^{11}$ GeV and (b) the so called
Gauge Mediated Susy Breaking (GMSB) type models\cite{dine2}, where SUSY breaking is mediated by 
the gauge interactions of the standard models via one and two loop
radiative corrections (for instance squark squared masses are of order of  $m^2_{\tilde{q}}\simeq 
\left({{\alpha}\over{4\pi}}\right)^2\Lambda^2$). So the natural scale
of SUSY breaking in these models is of order $10-100$ TeV raising the
possibility that they are accessible to low energy tests at current and
planned accelerators. The GMSB models have the extra advantage that
the FCNC effects are naturally suppressed 
due to the fact that at the scale $\Lambda$, the squark masses are
all degenerate due to the flavor blindness of the standard model gauge group.
However they suffer from the so called $\mu$ problem since the
gauge interactions being $U(1)_{PQ}$ symmetric do not generate a
$\mu$ term. One can however add new interactions to the model
to solve this problem\cite{dvali}.

In this paper, we discuss another class of models where the SUSY breaking
is caused by the existence of an anomalous local $U(1)$ around the Planck
scale, which due to the Green-Schwarz mechanism for anomaly cancellation
leads\cite{witten} to a linear $D$-term which leads
to supersymmetry breaking via the Fayet-Illiopoulos mechanism. Attempts
have recently been made\cite{dudas,pomarol,nakano,riotto} to build realistic
particle physics models using this new SUSY breaking mechanism.
This SUSY breaking
is fed down to the visible sector\cite{pomarol} both by the $D$-term as well as
by the supergravity effects. It was shown in Ref.\cite{pomarol} that in
the resulting theory, the gaugino masses are suppressed. It was also
conjectured in Ref.\cite{pomarol}
 that the FCNC and CP violating effects in these models are
suppressed. In Ref.\cite{riotto}, explicit models were constructed
where both the FCNC effects as well as the electric dipole moment of
the neutron were shown to naturally suppressed. It was further shown
how one may have a suppressed $\mu$ term in these models using the
Giudice-Masiero mechanism\cite{giudice} and how one may hope to
understand the fermion mass hierarchies. The primary features of these
models that helped in solving the FCNC and the SUSY CP problems are that 
the relative squark mass  difference (between the like-charged 
squarks of the first two generations) $\delta_{q}\equiv
\Delta m^2_{\tilde{q}}/m^2_{\tilde{q}}$, the gaugino masses
relative to the average squark masses $\delta_{\lambda}\equiv m_{\lambda}/
m_{\tilde{q}}$ as well as $\mu/m_{\tilde{q}}$ and $A/m_{\tilde{q}}$
are all small, with the suppression characterized by a  common 
parameter $\epsilon\simeq 10^{-2}$. The $\epsilon$ parameter is related
to the magnitude of the $U(1)$ anomaly\cite{witten2} which can be calculated
in terms of the low energy fermion spectrum and is therefore not an arbitrary
parameter.

It is the goal of this paper to elaborate on the various results of the
Ref.\cite{riotto} as well as to study the naturalness of the various
higher dimensional non-renormalizable terms necessary in this model in terms
of higher scale renormalizable theories. We find that indeed it is possible
to generate only the desirable non-renormalizable terms in the effective low
energy theory. We also study in more detail the implications of such
underlying theories for fermion masses as well as FCNC and CP violating
effects. The paper is organized as follows: in Sec. 2, we present the effective
Lagrangian for the for the model and discuss the electroweak symmetry breaking;
in Sec. 3, we discuss the electrweak symmetry breaking in the model;
in Sec. 4, we show how the suppressions of the FCNC and SUSY CP effects arise; 
in Sec. 5, the implications of the model
for the fermion masses is discussed. In Sec. 6, 
we discuss the naturalness of the low energy effective action in terms of
an underlying renormalizable theory. In Sec. 7, we discuss the cancellation
of the cosmological constant in the model and make some brief comments
on the mass spectrum of the theory.
                                    
\section{Effective action for the model}

As already alluded to, the crucial feature of the model is the
existence of a $U(1)$ gauge group, which is anomalous.
The $U(1)$ group may be assumed to emerge from string theories.
We will assume that the anomaly is cancelled by the Green-Schwarz
mechanism. Since the $U(1)$ is anomalous, i.e. ${\rm Tr}{\bf Q}\neq 0$, a
Fayet-Illiopoulos term which is a linear  $D$-term is always generated 
as a quantum effect. We further assume that  there is a pair 
of hidden sector fields denoted by $\phi_+$ and $\phi_-$ 
which have $U(1)$ charges
$+1$ and $-1$ respectively and that the fields of the standard model
also carry $U(1)$ charges. It is the assignment of the $U(1)$ charges
to quark superfields that help in the solution of the FCNC 
and CP problems and in qualitatively explaining the fermion mass hierarchy.
We assume the following $U(1)$ charge assignment for the fields of
the model (Table I):

\begin{center}
{\bf Table I}
\vskip 0.1cm

\begin{tabular}{|c||c||c||c||c||c||c|}  \hline
Fields & $\phi_+$ &$\phi_-$ &$ H_u$ &$ H_d$ & $Q_3,u^c_3,d^c_3$ &$ Q_i,
u^c_i,d^c_i$\\ \hline
$U(1)$-charge & +1 &-1& 0& +2 & 0 & +1 \\ \hline\end{tabular}

\end{center}

\noindent {\bf Table Caption:} The $U(1)$ quantum numbers of the various
fields in the theory.

\vspace{4mm}

In the above table, $i=1,2$ for the first two generations. We have omitted the
leptonic fields for simplicity; one could assign them same charges as the
down quark sector. 
Note that both the superpotential $W$ and the Kahler potential $K$ of the model 
must be invariant under the anomalous $U(1)$ symmetry. 
Let us first discuss the superpotential $W$, which we write as
$W= W_0 + W_1 +W_2+W_3$, 
where
\begin{eqnarray}
W_0&=&m\phi_+\phi_-,   \nonumber\\
W_1&=&h_u Q_3H_u u^c_3,   \nonumber\\
W_2 &=& (h_{u,3i}Q_3 H_u u^c_i)\frac{\phi_-}{M_{P\ell}}+
(h_{d,33} Q_3H_d d^c_3)\frac{\phi^2_-}{M^2_{P\ell}}
+(h_{u,ij}Q_iH_u u^c_j)\frac{\phi^2_-}{M^2_{P\ell}}\nonumber\\
&+&(h_{d,3i}Q_3H_dd^c_i)\frac{\phi^3_-}{M^3_{P\ell}} 
+(h_{d,ij}Q_iH_dd^c_j)\frac{\phi^4_-}{M^4_{P\ell}},\nonumber\\
W_3&=& (W_1+W_2)\frac{\phi_+\phi_-}{M^2_{P\ell}}
+\cdot\cdot\cdot.
\end{eqnarray}	
In the above equation, the ellipses denote all other higher dimensional
terms allowed by the gauge symmetry and make very
small contributions to the effects isolated below. 
The parameter $m$ is chosen to be of the order of the weak scale.

Let us now write down the Kahler potential $K(z_i,z_i^*)$ for 
the fields of the  model generically indicated by $z_i$. It  can
be written as the sum of two terms: one that involves the bilinear
terms of the form $z_i^*z_i$ and a second piece that involves mixed
terms which are strongly constrained by the $U(1)$ symmetry.
\begin{eqnarray}
K&=&K_0 +K_1, \nonumber\\
K_0&=&\sum_i |z_i|^2,\nonumber\\
K_1&=& \lambda H_uH_d\frac{\phi_-\phi^{\dagger}_+}{M^2_{P\ell}}+ 
{\rm h.c.} +\cdot\cdot\cdot.
\end{eqnarray}
In order to proceed further, we have to write down the potential
of the model involving the scalar fields $\phi_{\pm},~H^0_u,~H^0_d$ 
and determine the vacuum state. 
The part of the potential containing the $\phi_{-}$ and $\phi_{+}$ fields reads
\begin{eqnarray}
V &=& m^2(|\phi_+|^2+|\phi_-|^2)\nonumber\\
&+&\frac{g^2}{2}
\left({2}|H^0_d|^2+|\phi_+|^2-|\phi_-|^2+
\xi\right)^2.
\end{eqnarray}
where we have ignored the terms of order $m/M_{P\ell}$ or less.
Before discussing the minimization of the full potential, let us consider the
part of $V$ setting $H^0_u=H^0_d=0$. It is easy to see that its minimum
breaks supersymmetry as well as the anomalous $U(1)$ gauge symmetry 
with \cite{pomarol}
\begin{eqnarray}
\langle \phi_-\rangle &=&\left(\xi-{{m^2}\over{g^2}}\right)^{1/2},\:\:
\langle \phi_+\rangle = 0\\ \nonumber
\langle F_{\phi_+} \rangle &=&m\left(\xi-{{m^2}\over{g^2}}\right)^{1/2}.
\end{eqnarray}
If we parameterize $\xi=\epsilon M^2_{P\ell}$, for $m\ll M_{P\ell}$, we have
$\langle \phi_-\rangle\simeq \epsilon^{1/2}M_{P\ell}$ and $\langle F_{\phi_+}
\rangle \simeq \epsilon^{1/2}m M_{P\ell}$. Assuming that $\xi$-term is
induced by loop effects, one can estimate\cite{witten,pomarol} 
$\xi={{g^2{\rm Tr}{\bf Q}M^2_{P\ell}}\over{192\pi^2}}$, so that $\epsilon$ 
can be assumed to be of order $10^{-2}$. It was pointed out in 
ref.\cite{pomarol} that the gaugino masses are generated in this model
by superpotential terms of type 
$ \lambda'W^{\alpha}W_{\alpha}\left(\frac{\phi_+\phi_-}{M^2_{P\ell}}\right)$.
As a result, one gets gaugino masses to be 
$m_{\lambda_{g}}=\lambda'\epsilon m$. If we choose $m\simeq 1$ TeV, then
we need $\lambda'\sim 5$ to get the gluino mass at $M_Z$ of 100 GeV as
required by experiments.

From the $K_1$ term in the Kahler potential supergravity effects  induce a
$\mu$-term  by means of  the Giudice-Masiero mechanism \cite{giudice}. 
Indeed, $K_1$ induces at low energy  the operator
\begin{equation}
\lambda\int d^4\theta  H_uH_d\frac{\phi_-\phi^{\dagger}_+}{M^2_{P\ell}},
\end{equation}
giving rise to a $\mu$-term, with  $\mu=\lambda \epsilon m$. Notice 
that the corresponding $B$-term in the potential is induced at order 
$\epsilon^2$, by the term $H_uH_d\phi^2_-\phi_+\phi^{\dagger}_+/{M^4_{P\ell}}$.
There are bigger contributions to $B\mu$ from the renormalization
group running of the parameters from the Planck scale down.

\section{Electroweak symmetry breaking}

We integrate out the heavy field $\phi_{-}$ to obtain the 
effective potential of the light fields. Minimization with respect 
to  $\phi_{-}$ gives
\begin{equation}
|\phi_{-}|^2=\xi + |\phi_{+}|^2+{2}|H^0_d|^2-\frac{m^2}{g^2}.
\end{equation}
The effective potential of the fields $(\phi_{+},H^0_d,H^0_u)$ is 
at the leading order in $m^2/M^2_{P\ell}$
\begin{eqnarray}
V&=&2 m^2 |\phi_{+}|^2+m^2_{H_u}|H_u^0|^2+m^2_{H_d}|H_d^0|^2\nonumber\\
&-&m_3^2\left(H_u^0H_d^0+\:\:{\rm h.c.}\right)
+\:D{\rm -terms},\nonumber\\
m^2_{H_d}&=&|\mu|^2+{2}m^2+m_0^2,\nonumber\\
m^2_{H_u}&=&|\mu|^2+m_0^2,\nonumber\\
m^2_3&=&B\mu.
\end{eqnarray}
where we have indicated by "$D$-terms" the usual $D$-terms coming 
from $SU(2)\otimes U(1)$ and $m_i^2$ denotes the supersymmetry 
soft-breaking terms coming from supergravity, $m_0^2\sim \epsilon m^2$.
Note that all the values in the above equation are at the Planck scale.
They have to be extrapolated down to the weak scale, when we expect
that $m^2_{H_u}\simeq |\mu|^2+m^2_2$ with $m^2_2\leq 0$ so that
$H_u$ has a vacuum expectation value. However since $m^2_{H_d}$ is proportional to $m^2$
at the Planck scale, we expect it to remain sizable at the weak scale.
This implies that our model will prefer a large $\tan\beta$. Also
from the equation for eletroweak symmetry breaking:
\begin{eqnarray}
\frac{1}{2} M^2_Z=\frac{m^2_{H_d}-m^2_{H_u} \tan^2\beta}{\tan^2\beta-1}
\end{eqnarray}
we see that for $m_{H_d}\simeq 500$ GeV, a value of $\tan\beta\simeq 10$  
may be enough to get $M_Z$ of the desired order. But for instance
$\tan\beta\simeq 1$ is not at all adequate unless $m\simeq 100$ GeV, 
in which case we will get much too small a value for the gluino masses. 

Let us now look at other parameters of the theory. It is clear from the
the Eq. (1) that $A_u$ as well as $A_d$ suppressed by powers
of $\epsilon$ (Table II):
\begin{center}
{\bf Table II}
\vskip 0.1cm
\begin{tabular}{|c||c||c||c||c||c||c|} \hline
$A$ in units of $m$ & $A_{u,33}$ &$A_{u,3i}$&$A_{u,ij}$& $A_{d,33}$&
$A_{d,3i}$&$A_{d,ij}$\\ \hline 
$\epsilon$ suppression & $\epsilon$ &$\epsilon^{3/2}$&$\epsilon^2$&
$\epsilon^2$&$\epsilon^{5/2}$ &$\epsilon^3$\\ \hline
\end{tabular}
\end{center}

\noindent{\bf Table caption:} Degree of suppression of the various 
A-parameters in the theory.

\vspace{4mm}
 Note however that these are the values at the
Planck scale and they will evolve to higher values at the weak scale. It is
however important to note that both the values of $A$ and $B$ remain of
order $\epsilon$ at most since the value of $B$ at weak scale is proportional
to $ m_{\lambda_g}$ times the renormalization logarithm factor and similarly
for $A$. For instance a crude estimate would lead to 
$B\mu\simeq{\lambda^{\prime}}^2\epsilon^2 m^2$, which for $m\simeq 500$ GeV
can be of order $(50~GeV)^2$ or so, for $\epsilon\simeq 1/30$. 

\section{ Flavor Changing Neutral Current Effects and the electric dipole
moment of the neutron}

Let us now discuss the FCNC effects in this model. To study this, we note
that squark masses $m^2_{\tilde{q}}$ (both left and right handed types) 
receive two
contributions: a universal contribution from the $D$-term which is of 
order $m^2$ and a non-universal contribution from the supergravity Kahler
potential of order $ F^2_{\phi_+}/M^2_{P\ell}\equiv \epsilon m^2$. 
As both these 
contributions are extrapolated from the Planck scale down to the weak scale
the pattern of the first two generation squark masses remain practically
unchanged whereas the masses of the stop receive significant contributions.
It was noted in \cite{dine} that in order to satisfy the present observations
of FCNC effects (such as $K^0-\bar{K}^0$ mixing), the mixings between the
$\tilde{s}$ and the $\tilde{d}$ squarks (i.e. $m^2_{\tilde{s}\tilde{d}}$)
in the flavor basis or the squark mass differences 
between the first two generations in the mass basis
must satisfy a stringent constraint. In the flavor basis, it is given by
(see Dugan et al., in \cite{dine}), Im$\left({{m^4_{\tilde{s}\tilde{d}}}\over
{m^4_{\tilde{q}}}}\right)\leq 6\times 10^{-8}{{m^2_{\tilde{q}}}\over{m^2_W}}$.
We have assumed the phases in our model to be arbitrary; therefore the most
stringent constraint comes from the CP-violating part of the $K^0-\bar{K}^0$
mass matrix.
In our model, $m^2_{\tilde{s}\tilde{d}}$ arises purely from the supergravity
effects are of order $\sim \epsilon m^2$ and the above FCNC constraint
is satisfied if $\epsilon\simeq 10^{-2}$ or so. Thus our model
confirms the conjecture of Ref.\cite{pomarol}.

The electric dipole moment of the neutron $d^e_n$ 
in supersymmetric  models have been
discussed in several papers\cite{nedm} and it is by now well-known that
the gluino intermediate states in the loop graph contributing to the
$d^e_n$ gives a contribution which is some three orders of magnitude
larger than the present experimental upper limit
for generic values of the parameters. The situation is different in our
model since we see that a number of parameters of the model such as the
gluino masses, the $A$ and $B$ are down by powers of $\epsilon$. In order
to see the impact of this on the NEDM, we will again consider the charge
assignment for the first model 
where the Kahler potential induced mass splittings
in the squark masses are of order $\epsilon m^2$. For the gluino contribution,
we borrow from the calculation of Kizukuri and Oshimo \cite{nedm},
which gives:
\begin{eqnarray}
d^e_n&=& \frac{2e\alpha_s}{3\pi}\left( \sin\alpha_u A_u -
\sin\theta_\mu {\rm cot}\beta |\mu|\right)\nonumber\\
&\times&\frac{m_u}{m^2_{\tilde{q}}}\frac{1}{
m_{\lambda_3}}I\left(\frac{m^2_{\tilde{q}}}{m^2_{\lambda_3}}\right),
\end{eqnarray}
where $\alpha_u=\theta_{A_u}-\theta_{\lambda_3}$ is the differerence 
between the phases of the $A$-term and the gluino mass. 
 $m_{\tilde{q}}$ denotes the mass of the heavier of
the two eigenstates. Since in this model, $m_{\lambda_3}\simeq \sqrt{\epsilon}
m$ and $m_{\tilde{q}}\simeq m$, one 
finds that $I\simeq \epsilon$. This  leads to  
$d^e_n\simeq {{2\alpha_s}\over{3\pi}}\epsilon^{3/2} {{m_u}\over
{m^2}}$. Here we have used the fact that $A\sim \epsilon m$; 
$\mu \sim {\epsilon}m$. For $\epsilon \simeq 10^{-2}$, this 
gives an additional suppression of $10^{-3}$ over the prediction of generic
parameter values of the MSSM (i.e. even for $m\simeq 100$ GeV, we get
$d^e_n\simeq 10^{-25} $ e$\cdot$ cm).   
There is also a down quark contribution with a similar expression; but
in this model $A_d \ll A_u$, only the second term in the above equation
with $\cot\beta$ replaced by $\tan\beta$. By the same line of reasoning as
above, this term also also naturally suppressed.
We wish to point out that the above suppression depends on the fact
that $Q_1,u^c_1,d^c_1$ all have nonzero $U(1)$ charge. If on the other hand,
$d^c$ and $u^c$ had zero charge, their dominant mass would come
from the supergravity effect and, as a result,  
$m^2_{\tilde{d^c}}\sim m^2_{\tilde{u^c}}\simeq \epsilon m^2$. 
The  above gluino 
contribution to $d^e_n$ would then be less suppressed (by a factor
$\sqrt{\epsilon}$ rather than $\epsilon^{3/2}$). 
                                                       
\section{fermion mass matrices}
 Let us now discuss the pattern of fermion masses suggested by this
model. First note that only the Yukawa coupling $Q_3H_uu^c_3$ is allowed
without any suppression from the $\epsilon$ factor explaining why the
top quark has large mass \cite{jain,cohen}. On the other hand, the other
Yukawa couplings are suppressed with powers of $\epsilon$ qualitatively
explaining why their masses are so much smaller than the top quark mass.

From the superpotential in Eq.(1), we get the
following kind of up and down quark mass matrices.
\begin{eqnarray}
M_u=m_1\left(
\begin{array}{ccc}
\epsilon & \epsilon & \sqrt{\epsilon} \\
\epsilon & \epsilon & \sqrt{\epsilon} \\
\sqrt{\epsilon} & \sqrt{\epsilon} & 1
\end{array}\right)
\end{eqnarray}
and
\begin{eqnarray}
M_d=m_2
\left(\begin{array}{ccc}
\epsilon^{2} &\epsilon^{2} & \epsilon^{3/2} \\
\epsilon^{2} & \epsilon^{2} & \epsilon^{3/2} \\
\epsilon^{3/2} & \epsilon^{3/2} & \epsilon
\end{array}\right).
\end{eqnarray}
where $m_{1,2}$ are mass parameters related to the $v_{u,d}$ and the
Yukawa couplings. The first interesting prediction of this model is that
$m_c\simeq \epsilon m_t$ and $m_b\sim \epsilon m_t$ . Note that these
are in qualitative agreement with observations. $m_s\sim \epsilon^2 m_t$
may also be acceptable if $\epsilon$ is not literally $10^{-2}$ but somewhat
larger. Furthermore, if there is a horizontal symmetry between the first
and the second generation, then we expect $m_u\sim m_d\sim 0$ which is also
not unreasonable. We find this an encouraging aspect of the model that
needs further study beyond the scope of this paper.

\section{Underlying high scale theory and naturalness of the Kahler
and the superpotential}

In this Section we want to show that the superpotential as well as the 
Kahler potential chosen can indeed arise as an effective theory
from an underlying renormalizable model
which is valid around the Planck scale. The reason for such an exercise
is the following: note that we show that in our model the
$\mu$ term is suppressed naturally to the desired electroweak scale
because it arises from the Kahler potential term 
$H_uH_d\phi_-\phi^{\dagger}_+$. 
However, if we look naively at the model, the gauge symmetries 
also allow a superpotential term $\int d^2\theta
H_uH_d\phi^2_-/M_{P\ell}$ which would lead to a $\mu\simeq M_{P\ell}$.
This would of course be undesirable. We will show in this section there is an
underlying theory where only the first term arises as an effective term 
at low energies and not the latter. Similarly all the higher dimensional
superpotential terms that are responsible for the quark masses can also
arise in this theory. This makes our choice of the Kahler as well as
superpotential technically natural.

Let us assume that theory above the scale $\epsilon^{1/2}M_{P\ell}$ is
characterised by the following fields in addition to the
ones already given earlier: $SU(2)_L$ doublet vectorlike, colorless
fields: $L, \bar{L} $ and color singlet and $SU(2)_L$ singlet fields
$N^c_i$ with $i=1,2,3$ and color triplet or anti-triplet fields
$D^c,\bar{D^c}, D^{c\prime},\bar{D^{c\prime}}$. These particles
are assumed to have masses $\sim M_{P\ell}$ and are
expected to decouple below $M_{P\ell}$ so that at
$\mu\sim \epsilon^{1/2}M_{P\ell}$, the theory will have the same
structure as in sec.II. Naive decoupling arguments would seem to
support this assumption.
The $U(1)$ charge assignment for these fields are given in Table III.
\begin{center}
{\bf Table III}
\vskip 0.1cm
\begin{tabular}{|c||c||c||c||c||c|}\hline
Fields & $L,N^c_3, \bar{D^{c\prime}}$ &$ \bar{L}, N^c_1,D^{c\prime}$ 
&$N^c_2$ & $\bar{D^c}$ & $\bar{D^c}$\\ 
\hline
$U(1)$-charge & +1 & -1& 0 & +2 & -2\\ \hline
\end{tabular}

\noindent{\bf Table caption:} The $U(1)$ charge assignment of the fields
of the underlying theory.

\vspace{4mm}
\end{center}
We also assume that there is a $Z_2$ symmetry under which the fields
$L,\bar{L}, N^c_i$ ($i=1,2,3$) are odd and the remaining fields are
even. The allowed gauge and $Z_2$ 
invariant couplings involving the heavy and light
fields can be written as a superpotential $W_5$
\begin{eqnarray}
W_5&=& H_uL_1N^c_1 + M_1 L\bar{L}+M_2 N^c_1N^c_3 +M_3 N^c_2N^c_2 \\ \nonumber
&+& N^c_1N^c_2\phi_++ N^c_2N^c_3\phi_-+H_dN^c_1\bar{L} \\ \nonumber
&+&Q_3H_dD^c + M_D D^c\bar{D^c} +\bar{D^c}\phi_-D^{c\prime} + 
M_{D'}D^{c\prime}\bar{D^{c\prime}}+\bar{D^{c\prime}}\phi_-d^c_3
\end{eqnarray}

It is the easy to see that $\mu$, $B\mu$ and $Q_3H_dd^c_3\phi^2_-$ terms are
generated by the diagrams in Fig. 1, 2 and 3 respectively. 
On the other hand a term
of the form $H_uH_d\phi^2_-$ is never generated in the effective low energy
theory. It is possible to add to the theory extra $D^c$ and $\bar{D^c}$
type fields with
appropriate quantum numbers so that the other higher dimensional terms that
lead to quark masses for lower generations can emerge. 

One may have hoped that this underlying theory could be used to completely
 eliminate the R-parity violation from the effective 
low energy theory in a natural manner. 
It however turns out that in this particular example, it does not happen.
There are however suppressions by powers of $\epsilon$ in front of the 
various R-parity violating couplings.

\section{Cosmological constant and sparticle spectrum}

The model chosen so far has a cosmological constant of order $V_0\sim
\epsilon m^2 M^2_{P\ell}$. It is however easy to set it
to zero by adding to the superpotential of the model (Eq. (1))
a constant term denoted by $\beta^3$, where $\beta$ has dimension of mass.
Requiring the cosmological constant to vanish implies that
\begin{eqnarray}
\beta^6=\frac{\epsilon m^2 M^4_{P\ell}-g^{-2}m^4M^2_{P\ell}}{3-\epsilon
+g^{-2}m^2 M^{-2}_{P\ell}}
\end{eqnarray}

The squark mass splittings in this case become of order
$\Delta m^2_{\tilde{Q}}\sim \frac{\epsilon}{3}m^2 $. This implies
that for $\epsilon\sim 1/30$, we get a suppression in the squark mass
splittings of order $10^{-2}$. The
gravitino mass can be estimated to be $m_{3/2}\sim
{(\frac{\epsilon}{3})}^{1/2}m$. 
                       
Let us also briefly comment on the expected masses of the superpartners
in this model. The two key parameters are the superpotential mass
$m$ and the anomaly factor $\epsilon$. We will assume $m\simeq
500-1000$ GeV. The value of $\epsilon$ is taken to be $\frac{1}{30}$.
We then find that at $M_{P\ell}$, the gluino mass $M_{\tilde{G}}\simeq
\lambda'\times(17-34)$ GeV. At the scale $M_Z$, one gets $M_{\tilde{G}}
\simeq \lambda'\times (51-68)$ GeV. So if chose $\lambda'\geq 2$, we
would be in compliance with the present experimental constraints. 
If on the other hand we chose $m=100$ GeV as an extreme example,
one would be driven to the light gluino scenario (which though not
favored, is perhaps not excluded \cite{farrar}). It thus appears
that the gluino mass could be a potential embarrasment for these models
if either the light gluino is definitively ruled out or one is unable to
add a new source for the gluino mass to the model. At the moment
we find these models to have so many attractive features that we wish
to pursue them as serious candidates hoping that this issue will
find a resolution. As far
as the chargino masses are concerned, if the corresponding $\lambda'$
is also chosen to be around two, the charginos states appear nearly
degenerate since the $\mu$-term in this model is also likely to be
small. A detailed investigation of the expected sparticle spectra
for plausible parameter ranges of the theory is presently under way.
   
Another point that distinguishes these models from the gauge mediated SUSY
breaking scenarios is that we expect the squarks of the first and
the second generation and the sleptons (assuming the leptons have
the same charge +1 as the corresponding quarks) to have nearly the same
mass. Note that in the GMSB models the sleptons are considerably lighter.

\section{Conclusion}

In conclusion, we have studied ways 
to construct interesting realistic supersymmetric models 
of quarks and leptons using
the idea that an anomalous $U(1)$ gauge symmetry is responsible for
generating supersymmetry breakdown. These models have the 
attractive feature that they solve
several fine tuning problems of the MSSM associated with FCNC effects and
electric dipole moment of the neutron. They also give desirable values for
the $A$, the $B$ and the $\mu$ parameters 
and also have the potential to qualitatively
explain fermion mass hierarchies. We also show how the effective higher
dimensional terms used in making the model realistic can emerge from
an underlying renormalizable theory in a natural manner.

\vskip 0.1 cm

The work of R. N. M. is supported by the NSF grant no. PHY-9421385
and the work of A. R. is supported by the DOE and NASA under Grant NAG5--2788.
R. N. M. would like to thank B. Brahmachari for discussions and A. R.
would like to thank E. Poppitz, S. Trivedi and E. Dudas for comments and
discussions.

\noindent {\bf Figure caption}

\noindent {\bf Figure 1} The one-loop diagram that leads to the effective
operator $H_uH_d\phi_-\phi^{\dagger}_+$ that gives the $\mu$-term at low
energies.

\noindent {\bf Figure 2} The one-loop diagram that leads to the $B\mu$ term
at low energies.

\noindent {\bf Figure 3} The tree diagram that gives the operator 
$QH_dd^c\phi^2_-$ at low energies.


\begin{references}
\begin{enumerate}


\bibitem{nilles} For reviews, see 
H.P. Nilles, Phys. Rep. {\bf 110}, 1 (1984); H. Haber
and G. Kane, Phys. Rep. {\bf 117}, 76 (1984); R. Arnowitt, A. Chamsheddine
and P. Nath, {\it Applied N=1 Supergravity}, World Scientific, 
Singapore (1984).

\bibitem{dine} M. Dugan, B. Grinstein and L. Hall, Nucl. Phys. {\bf B255},
413 (1985);  M. Dine, A. Kagan and R. Leigh, Phys. Rev. {\bf D48}, 4269
(1993); F. Gabbiani, E. Gabrielli, A. Masiero and L. Silvestrini,
hep-ph/9604387; A. Pomarol and D. Tommasini, Nucl. Phys. {\bf B466}, 3 (1996);
R. Barbieri and L. Hall, LBL-38381 (1996), hep-ph/9605224.

\bibitem{dine2}  M.~Dine, A.E.~Nelson, and Y.~Shirman,
Phys. Rev. {\bf D51} 1362 (1995) ;
M.~Dine, A.E.~Nelson, Y.~Nir, and Y.~Shirman, 
Phys. Rev. {\bf D53}, 2658 (1996) ;
M. Dine, hep-ph/9607294; M. Dine, Y. Nir and Y. Shirman, SCIPP 96/30 preprint,
hep-ph/9607397.

\bibitem{dvali} G. Dvali, G. Giudice and A. Pomarol, CERN preprint
hep-ph/9603238.

\bibitem{witten} M. Dine, N. Seiberg and E. Witten, Nucl. Phys. {\bf B289},
585 (1987); J. Atick, L. Dixon and A. Sen, Nucl. Phys. {\bf B292}, 109 (1987).

\bibitem{dudas} P. Binetruy and E. Dudas, hep-th/9607172.

\bibitem{pomarol} G. Dvali and A. Pomarol, CERN-TH/96-192, hep-ph/9607383;
G. Dvali and S. Pokoroski, hep-ph/9610431.

\bibitem{nakano} H. Nakano, hep-th/9404033; E. Dudas, S. Pokorski and C. Savoy,
Phys. Lett. {\bf B369}, 255 (1996); Y. Kawamura and T. Kobayashi, Phys. Lett.
{\bf B375}, 141 (1996).

\bibitem{riotto} R. N. Mohapatra and A. Riotto, hep-ph/9608441.

\bibitem{giudice} G. Giudice and A. Masiero, Phys. Lett. {\bf 206B}, 480
(1988).

\bibitem{witten2} E. Witten, Nucl. Phys. {\bf B188}, 513 (1981); W. Fischler
et al. Phys. Rev. Lett. {\bf 47}, 657 (1981).

\bibitem{nedm} J. Ellis, S. Ferrara and D. Nanopoulos, Phys. Lett. {\bf 114B},
 231 (1982); S. P. Chia and S. Nandi, {\it ibid }{\bf 117B}, 45 (1982);
J. M. Gerard et al. {\it ibid} {\bf 140B}, 349 (1984); M. Dugan, B. Grinstein
and L. Hall, Nucl. Phys. {\bf B255}, 413 (1985); Y. Kizukuri and N. Oshimo, 
Phys. Rev. {\bf D46}, 3025 (1992); R. Garisto, Nucl. Phys. {\bf B419}, 
279 (1994); Phys. Rev. {\bf D49}, 4280 (1994). 

\bibitem{jain} This method of understanding mass hierarchies is similar
in spirit to the papers: J. Lopez and D. Nanopoulos, Nucl. Phys. {\bf B338},
73 (1990); E. Papargeorgio, Phys. Lett. {\bf B343}, 263
(1995); L. Ibanez and G. G. Ross, {\it ibid} {\bf B332}, 100 (1994);
V. Jain and R. Shrock, {\it ibid } {\bf B232}, 83 (1995); P. Binetruy
and P. Ramond, {\it ibid} {\bf B350}, 49 (1995), P. Binetruy, S. Lavignac,
S. T. Petcov and P. Ramond, hep-ph/9610481.

\bibitem{cohen} A. Cohen, D. Kaplan and A. Nelson, hep-ph/9607394.

\bibitem{farrar} G. R. Farrar, Phys. Rev. {\bf D51}, 3904 (1995);
L. Clavelli, Phys. Rev. {\bf D46}, 2112 (1992).

\end{enumerate}
\end{references}
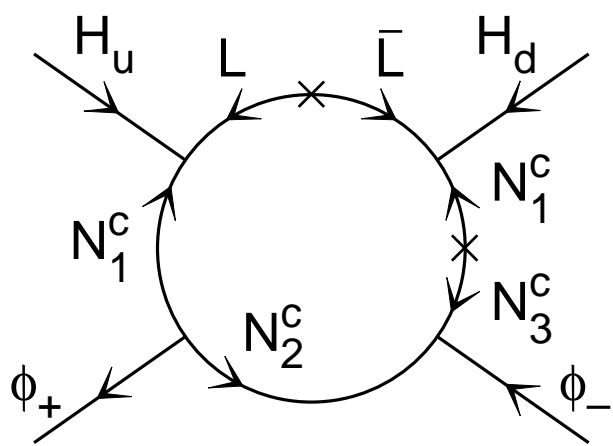
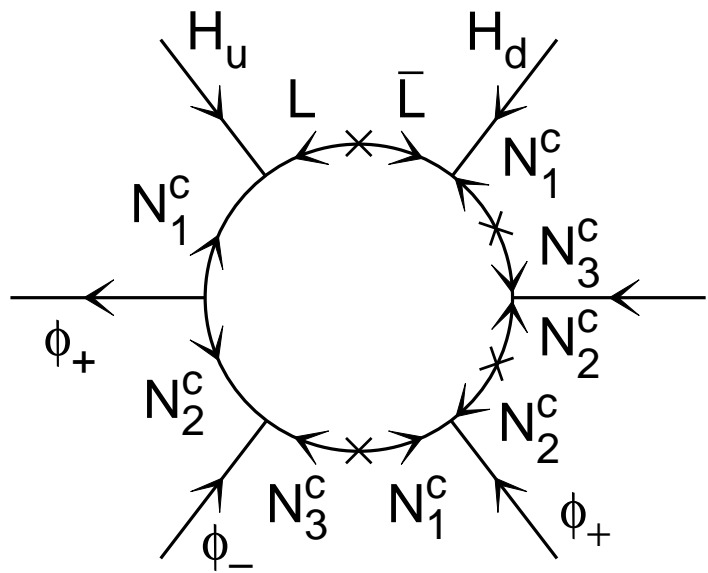
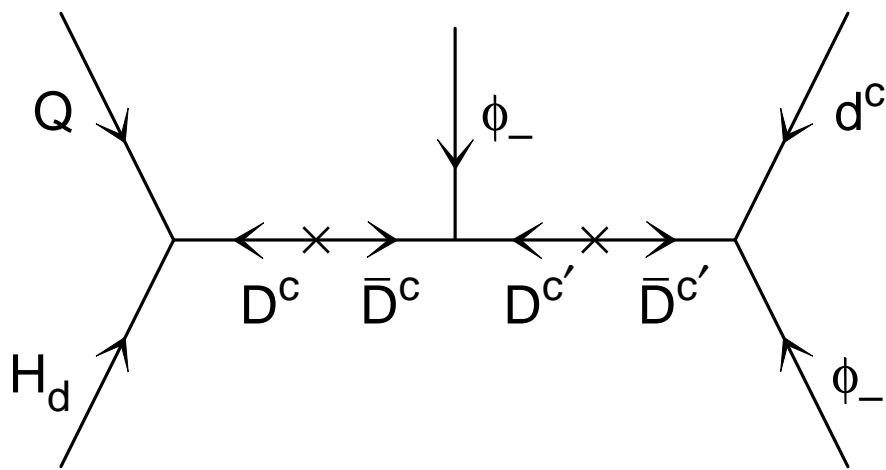
\end{document}